\documentclass[10pt,prb,twocolumn,showpacs]{revtex4}
\usepackage{graphicx}
\usepackage{epsf}
\usepackage{ifthen}
\usepackage{amsmath}
\usepackage{amssymb}
\usepackage{subfigure}
\usepackage{psfrag}
\usepackage{float}
\usepackage{subeqnarray}
\begin{document}
\title{Vibrational properties of nano-graphene}
\author{Sandeep Kumar Singh}
\affiliation{Department of Physics, University of Antwerp, Groenenborgerlaan 171, B-2020 Antwerpen, Belgium}
\author{F. M. Peeters}
\affiliation{Department of Physics, University of Antwerp, Groenenborgerlaan 171, B-2020 Antwerpen, Belgium}
\date{\today}
%Abstract

\begin{abstract}
The eigenmodes and the vibrational density of states of the ground state configuration of graphene clusters are calculated using atomistic simulations. The modified Brenner potential is used to describe the carbon-carbon interaction and carbon-hydrogen interaction in case of H-passivated edges. For a given configuration of the C-atoms the eigenvectors and eigenfrequencies of the normal modes are obtained after diagonalisation  of the dynamical matrix whose elements are the second derivative of the potential energy. The compressional and shear properties are obtained from the divergence and rotation of the velocity field. For symmetric  and defective clusters with pentagon arrangement on the edge, the highest frequency modes are shear modes. The specific heat of the clusters is also calculated within the harmonic approximation and the convergence to the result for bulk graphene is investigated.

\end{abstract}
\pacs{63.22.Kn, 63.22.Rc, 65.80.Ck}

\maketitle

\begin{section}{Introduction}
  During the past few years considerable attention has been paid to the study of the spectral properties such as the energy spectrum,
 the eigenmodes (i.e. phonon spectrum), and the phonon density of states of carbon
 allotropes such as graphene~\cite{Zimmermann, Maultzsch, Jian, Gillen}, carbon nanotubes~\cite{Gao, Saito}, fullerenes~\cite{Menon,Barszcz, Malolepsza}, graphene flakes~\cite{Bera} and carbon clusters~\cite{Agrawal,Breda}.

 Raman spectroscopy is a widely used tool to study the vibrational modes of carbon based materials~\cite{Dresselhaus, novoselov, gaim}. Because the normal modes of  molecules are unique, they have their  own set
 of Raman frequencies. Similarly for graphene clusters their  phonon modes will give us information on e.g. their structural and dynamical properties.

 Lattice dynamics calculations of a single graphene sheet were done by Rao et al.~\cite{Rao}  based on C-C force constants~\cite{venktaraman, venktaraman1} which were obtained from a fit to the two dimensional, experimental phonon dispersion. Good agreement with  Raman spectra of single wall carbon nanotubes was found. Tommasini et al.~\cite{Tommasini} presented a semiempirical approach for modeling the off-resonance Raman scattering of molecular models of confined graphene and compared with the results from density functional theory calculations. Pimenta et al.~\cite{Pimenta} reviewed the defect-induced Raman spectra of graphitic materials  and presented Raman results on nanographites and graphenes. Ryabenko et al.~\cite{Ryabenko} analyzed the spectral properties of
 single-wall carbon nanotubes encapsulating fullerenes and found that  fullerene
 alters the average diameter of the  electron cloud around the single
 wall carbon nanotube. Michel et al.~\cite{Michel} presented a unified theory
 of the phonon dispersion and elastic properties of graphene, graphite and graphene multilayer systems. They started from a fifth-nearest-neighbor force-constant model derived from the full in-plane phonon
 dispersions of graphite~\cite{Mohr}, and extended this model with interplanar interactions,  and found that  the graphite eigenfrequency $\omega_{B_{2g_{1}}} \approx$ 127 cm$^{-1}$ is reached within a few percent for N=10 layers which agrees with similar results obtained for the electron spectrum~\cite{Partoens}. The structural, dynamical and thermodynamical properties of carbon allotropes were computed by Mounet et al.~\cite{Mounet} using a combination of density-functional theory total-energy calculations and  density-functional perturbation theory lattice dynamics in the generalized gradient approximation. Very good agreement was found for the structural properties and the phonon dispersion.

High-level ab initio calculations were carried out to reexamine the relative stability of bowl, cage, and ring isomers of $C_{20}$ and $C_{20}^{-}$ by An et al.~\cite{Wei} The total electronic energies of the three isomers showed a different energy ordering which depends on the used hybrid functional.

Finite size planar structures which are close to the ground state, and in particular nanographene-like structures were studied by Kosimov et al.~\cite{kosi} using energy minimization with the conjugate gradient (CG) method. The lowest energy, i.e. the ground state~\cite{kosi,kosi1} configurations were determined for up to N=55 carbon atoms.

 Here we will investigate the normal modes for the ground state configuration of flat carbon clusters, also called nanographene as function of the number of C atoms in the cluster. The content of the paper is as follows. First (see Sec.~2), we present the main computational approach that is used to calculate the normal modes. The normal modes for linear (N $\leq$ 5) and ring ($6 \leq N \leq 18$) clusters which are exactly one and two dimensional, respectively are discussed in Sec.~3. Next (see Sec.~4), the phonon spectrum of two dimensional nanographene (19$\leq$ N $\leq$ 55) and trigonal-shaped nanodisk~\cite{Ezawa,Palacios} as well as the effect of hydrogenation of the edge carbon atoms on the phonon spectrum are investigated. In Sec.~5, we calculate the phonon density of states of various clusters. By increasing the number of C atoms in the cluster we will show how the phonon spectrum of graphene appears.  We also calculate the temperature dependence of the  specific heat within the harmonic approximation in Sec.~6. Summary (see Sec.~7) close the paper.
\end{section}
%\clearpage
\begin{section}{Simulation Method}
The Brenner second generation reactive empirical bond order (REBO) potential function between carbon atoms is used in the present work. The values for all the parameters used in our calculation for the Brenner potential can be found in Ref.~\onlinecite{bren2} and are therefore not listed here. The Brenner potential (REBO) energy is given by:

\begin{equation}
{E_b}=\sum_{i} \sum_{j(>i)} [V^R(r_{ij})-b_{ij}V^A(r_{ij})].
\end{equation}

\noindent Here ${E_b}$ is the total binding energy, $V^R(r_{ij})$ and $V^A(r_{ij})$ are a repulsive and an attractive term, respectively, where ${r_{ij}}$ is the distance between atoms $i$ and $j$, given by

\begin{equation}
V^R(r)=f^C(r)(1+Q/r)Ae^{-\alpha r},
\end{equation}

\begin{equation}
V^A(r)=f^C(r)\sum_{n=1}^{3} B_n e^{-\beta_{n} r},
\end{equation}
where the cut-off function $f^C(r)$ is taken from the switching cutoff scheme

\begin{equation}
\label{hj}
\displaystyle{f_{ij}^{C}(r)=\left\{\begin{array}{l}
                         1  ~~~~~~~~~~~~~~~~~~~~~~~~~~~~~~~~~~~ r<D_{ij}^{min}\\
                        \big[1+cos\big(\frac{(r-D_{ij}^{min})}{(D_{ij}^{max}-D_{ij}^{min})}\big)\big] /2~~D_{ij}^{min}<r<D_{ij}^{max}\\
                         0 ~~~~~~~~~~~~~~~~~~~~~~~~~~~~~~~~~~~~r>D_{ij}^{max}
               \end{array}
\right.
}
\end{equation}
where $D_{ij}^{max}~- D_{ij}^{min}$ defines the distance over which the function goes from one to zero and $ A, Q, \alpha, B_n, \beta_n, (1\leq n \leq 3)$ are parameters for the carbon-carbon pair terms. Here $n$ is the type of bond (i.e. single, double and triple bonds).

The empirical bond order function used in this work is written as a sum of terms:

\begin{equation}
b_{ij}=\frac{1}{2}[b_{ij}^{\sigma-\pi}+b_{ji}^{\sigma-\pi}]+b_{ij}^{\pi},
\end{equation}
where the functions $b_{ij}^{\sigma-\pi}$ and $b_{ji}^{\sigma-\pi}$ depend on the local position and bond angles determined from their arrangement around each atom ($i$ and $j$, respectively) and is governed by the hybridization of the orbitals around the atom. The first term in Eq. (5) is given by,

\begin{equation}
b_{ij}^{\sigma-\pi}=\big[1+\sum_{k(\ne i,j)} f_{ik}^C(r_{ik})G(cos(\theta_{ijk}))\big]^{-1/2}.
\end{equation}
Here the angular function $G(cos(\theta_{ijk}))$ modulates the contribution of each nearest neighbour and is determined by the cosine of the angle of the bonds between the atoms $i-j-k$.

  Using molecular dynamic (MD) simulations with the Berendsen after corrector thermostat,  we obtain the equilibrium configuration of stable quasi-planer clusters. We found that about $10^{4}$-$5\times10^{5}$  MD steps  are needed to obtain an accuracy  of $10^{-5}$-$10^{-6}$ eV in the energy per particle. We used free boundary conditions in our simulation.

Here, we calculate the dynamical matrix numerically using the finite-displacement method. We displace  particle $j$ in the direction $\beta$ by a small amount $\pm u$, and evaluate the forces on every particle in the system $F_{\alpha,i}^{\pm}$ (we took typically u=0.003 ${\r{A}}$).
Then, we compute numerically the derivative using the central-difference formula
\begin{equation}
\frac{dF_{\alpha,i}}{du_{\beta,j}} =\lim_{u\to 0}\frac{F_{\alpha,i}^{+} - F_{\alpha,i}^-}{2u_{\beta,j}} = \frac{d^2 E_b}{du_{\alpha,i} du_{\beta,j}}.
\end{equation}

 Diagonalization of the dynamical matrix yields the eigenvalues and the eigenvectors from which we can calculate the vibrational modes.
 The eigenfrequencies are the square root of the eigenvalues. The
stability of the  configuration can be verified because all eigenvalues have to be real.

\begin{figure}[htb]
%  \begin{center}
\centering
    \includegraphics[width=10cm]{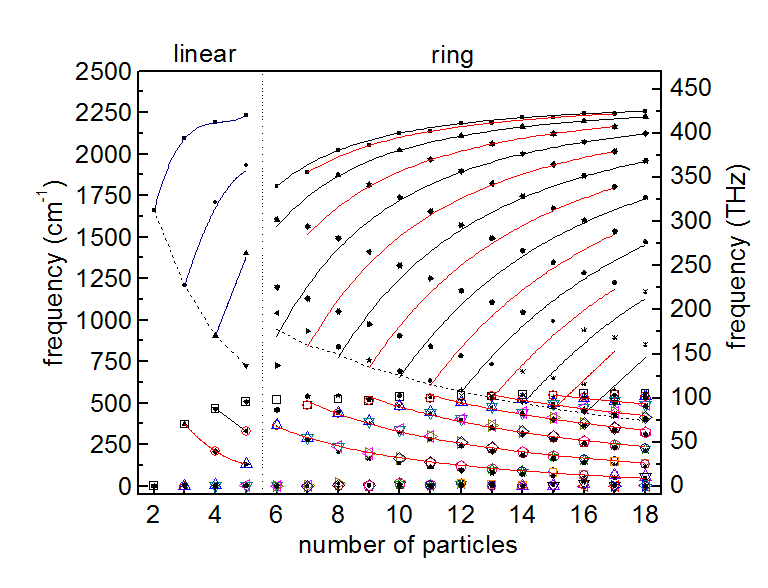}
%  \end{center}
  \caption{ (Color online)
 Spectrum of normal modes as a function of the number of particles in the cluster for linear N $\leq$ 5 and ring $6 \leq N \leq 18$ clusters. The solid curves are the analytical results from a simple chain (N $\leq$ 5) and ring ($6 \leq N \leq 18$) model. The dashed curve indicates the breathing mode. The symbols are the numerical results from the present simulation where the solid (hollow) symbols are the frequencies with corresponding eigenvectors which are in plane (out of plane). The frequencies that decrease with the number of particles are fitted to a $1/N$ dependence (solid curves).
}\label{fig1}
\end{figure}

\begin{figure}[htb]
\centering
%  \begin{center}
    \includegraphics[width=8.6cm]{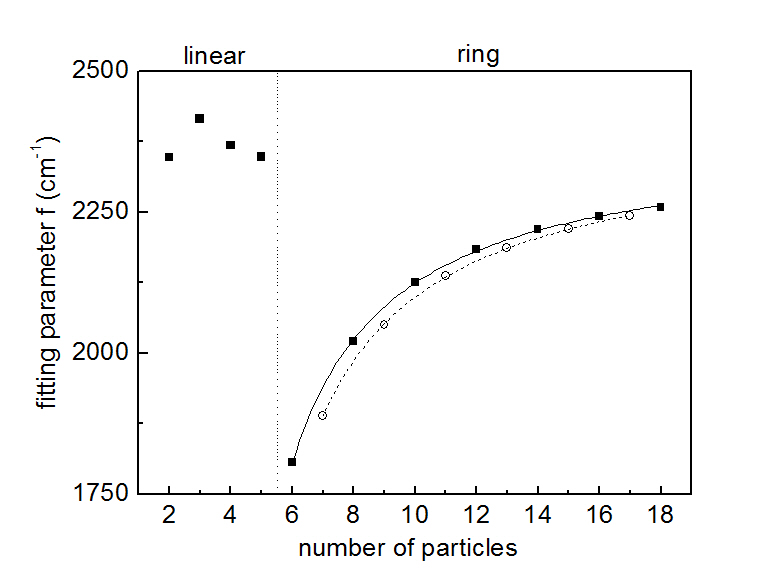}
%  \end{center}
  \caption{
Fitting parameter $f$ of linear chain (N $\leq$ 5) and ring ($6 \leq N \leq 18$) model of particles connected by identical springs as a function of the number of particles. For $N\geq6$ the results are fitted by a simple analytical expression  for even (full curve) and odd (dashed curve) clusters.
}\label{fitting}
\end{figure}
\end{section}
\begin{section}{LINEAR AND RING CLUSTERS}

The  energy of the normal modes corresponding to the ground-state configuration for linear and ring clusters is shown in  Fig.~\ref{fig1} as
  function of the number of C atoms in the clusters. The number of normal modes for N-particles moving in 3-dimensions is 3N. Some of these modes will be degenerate. There are 5 modes with $\omega=0$ which correspond to 3 uniform translations and 2 independent rotations of the whole cluster, due to the translational and rotational invariance of the cluster.

When clusters are exactly one dimensional, the
phonon modes correspond to oscillations along the chain or orthogonal to it. The normal modes of such a chain of N particles connected by springs with spring constant $\kappa$ and mass m can be calculated analytically \cite{James} and is given by $\omega=f \times \frac{sin(\phi_{n}/2)}{sin\phi_{N}}$ where
$f=2\sqrt{\frac{\kappa}{m}}$, and $\phi_{n}=(n-1)\pi/N$  $(n=1,2,\cdots,N)$.
For each value of N, we determined $f$ by taking the maximum frequency obtained from our model equal to our numerically found result. The obtained frequencies of our linear spring chain model (solid curves in Fig.~\ref{fig1}) agree remarkably well with our numerical results. For $N\leq5$, the obtained spring constant is plotted in Fig.~\ref{fitting} and is almost constant (less than 4$\%$ variation). As shown in Ref.~\cite{kosi1} all the C-C bonds in the chain are not identical and therefore all the spring constants between the C-atoms are  expected not to be identical, which is the reason why our simple model does not give perfect agreement with our numerical data.

Notice that the low frequency eigenmodes are not described by this simple model. Some of them correspond  to out-of-plane motion.
For cluster
N=2 with mode k=6 (k counts the eigenvalues in increasing order) and
for N=4 with mode k=12, the particles move in opposite direction to
each other along the chain as depicted in Fig.~\ref{linear}. While the cluster N=3 with mode k=9 and the cluster  N=5
with k=15 have alternating particles moving in opposite direction which makes an angle with the chain direction (see Fig.~\ref{linear}).

The breathing mode for the linear chain is shown by the dotted line  in Fig.~\ref{fig1}. For N=2 the excitation corresponding to the breathing mode is shown in Fig.~\ref{linear}(a). But for N=3, this mode is similar for N=2 but where the direction of oscillation makes an angle with the chain direction and the middle particle does not move due to conservation of momentum. Similarly, for N=5 we found that the breathing mode is similar to the one for N=4 except that the deflection  makes an angle with the  chain direction and  the middle particle is at rest (see Figs.~\ref{linear_breath}(a, b)). For mode k=6 and 7 with N=3,4,5 particles, the  particles on the boundaries of the clusters move in the same direction while the remaining particles move in the opposite direction with different amplitudes to conserve total angular momentum (see Fig.~\ref{linear_breath}(c)) for both in-plane and out-of-plane direction. The corresponding frequency of these modes can be reasonably well fitted by $\frac{a}{N}$+b (solid curve in left panel of Fig. 1) where the fitting parameters are (a,b)= (1809 $(\pm 103)$,  -234 $(\pm 27))$ cm$^{-1}$. And for mode k=8 and 9 with N=4, 5 particles, adjacent particles move in opposite direction with different amplitudes where the  central particle is at rest in the case of N=5 particles (see Fig.~\ref{linear_breath}(d)) for both in plane and out of plane direction and they can be fitted by the same function with (a,b)= (1783 $(\pm 103)$, 0.0) cm$^{-1}$.
\begin{figure}[htb]
\centering
%  \begin{center}
    \includegraphics[width=8.6cm]{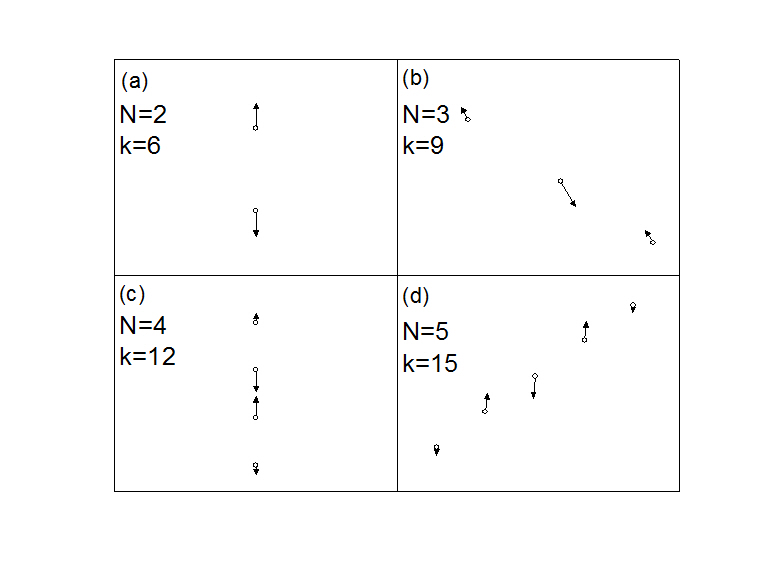}
%  \end{center}

\caption{ Eigenvectors of the highest frequency modes for the clusters with  N=2,3,4 and 5 particles for different mode number with frequency (a) $\omega_{6} = 1660$ cm$^{-1}$, (b) $\omega_{9} = 2092$ cm$^{-1}$, (c) $\omega_{12} = 2188$ cm$^{-1}$ and (d) $\omega_{15} = 2233$ cm$^{-1}$.}
\label{linear}
\end{figure}

\begin{figure}[htb]
\centering
%  \begin{center}
    \includegraphics[width=8.6cm]{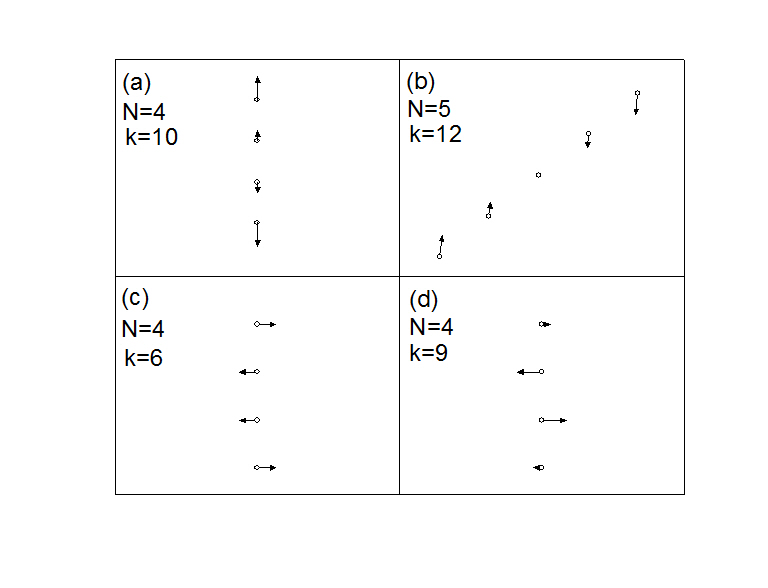}
%  \end{center}

\caption{ Eigenvectors  for the clusters with  N=4 and 5 particles for different mode number with frequency (a) $\omega_{10} = 905$ cm$^{-1}$, (b) $\omega_{12} = 725$ cm$^{-1}$, (c) $\omega_{6} = 150$ cm$^{-1}$ and (d) $\omega_{9} = 466$ cm$^{-1}$.}
\label{linear_breath}
\end{figure}
The ground state configuration for $C_{N}$ (N=6-18) consists of particles arranged in a ring.
For clusters that are exactly two dimensional, as is the case of these ring structures,
the phonon modes correspond to motion  in the plane or vertical to the plane.
There is no coupling between the in-plane and the out-of-plane motion. We can construct a simple model of equidistant  particles arranged in a ring, where neighbor particles are connected by a spring. This model can be viewed as the linear chain model with periodic boundaries. For fixed radius of the ring this results in the following
eigenfrequencies \cite{Jon} $\omega=f \times sin(k/2)$, where $k=2\pi n/N,$ $(n=1,2,\cdots,N)$. As before we determined $f$ such that the maximum frequency obtained from the model equals the numerical result. This fitted value is shown in Fig.~\ref{fitting} which can be described approximately by the function $f(N)=a \times \frac{(1+bN)}{(1+cN)}$, where: a=3200 cm$^{-1}$ and i) for even N clusters
b =  -0.295 ($\pm$ 0.001), and
c = -0.395 ($\pm$ 0.002); and ii) for odd N clusters
b =  -0.276 ($\pm$ 0.001), and
c = -0.369 ($\pm$ 0.001). Several of the numerical frequencies are reasonably well described by this simple model. Notice that the eigenfrequencies exhibit a clear discontinuous behavior when a cluster changes from a linear into a ring configuration. As an example we show in Fig.~\ref{18k} some of the interesting eigenmodes of a cluster with N=18 particles which
are arranged in a circle and is therefore pure two dimensional. The mode k=23 is the breathing mode while mode k=38 corresponds to radial out of phase radial oscillation of nearest neighbor particles. The modes k=53, and 54 correspond to pure angular oscillations of the particles.

\begin{figure}[htb]
\centering
%  \begin{center}
    \includegraphics[width=8.6cm]{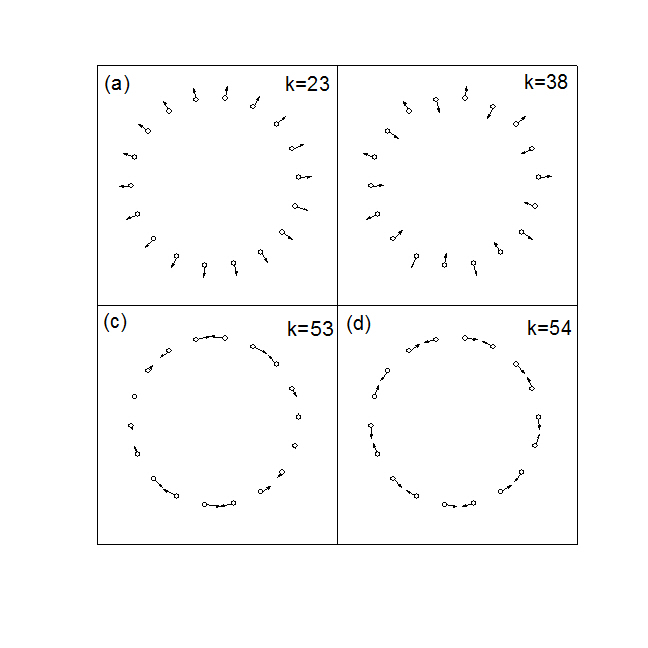}
%  \end{center}
  \caption{
Eigenvectors for a cluster with N=18 particles for different mode number with frequency (a) $\omega_{23} = 398$ cm$^{-1}$, (b) $\omega_{38} = 550$ cm$^{-1}$, (c) $\omega_{53} = 2224$ cm$^{-1}$ and (d) $\omega_{54} = 2259$ cm$^{-1}$.
}\label{18k}
\end{figure}

It is interesting to note that for an even number of particles,
there is a nearly common frequency (i.e. N=8 (545 cm$^{-1}$),
N=10 (543 cm$^{-1}$), N=12 (546 cm$^{-1}$), N=14 (547
cm$^{-1}$), N=16 (549 cm$^{-1}$), N=18 (550 cm$^{-1}$)) which
is almost independent of N (see Fig.~\ref{fig1}). This mode corresponds to out of phase oscillations of nearest-neighbor atoms (see Fig.~\ref{18k}(b) the k=38 mode for N=18 and Fig.~\ref{12-13}(a)
the k=24 mode for N=12). But for odd ring clusters i.e. N=13, the corresponding normal mode corresponds to  nearest-neighbor atoms oscillating  out of phase with different amplitudes except for the two neighbor  atoms that oscillate in phase (see Fig.~\ref{12-13}(b) k=23 (471 $cm^{-1}$)) which results in a lower frequency by about 20 \% as compared to the corresponding even N ring clusters.

    A  breathing mode exists for all the ring clusters as shown in Fig.~\ref{12-13}. The frequencies for N=8, 12, 13, 16, and 18 are 837 cm$^{-1}$, 585 cm$^{-1}$, 543 cm$^{-1}$, 447 cm$^{-1}$,  and 398.25 cm$^{-1}$, respectively, and have a clear dependence on N. The simple spring model \cite{zhou} predicts that this breathing mode has the frequency $\omega$=$(\pi \times f)/N$ which remarkably agrees with our numerical results.

The eigenmodes corresponding to  $\omega_{max}$ have  similar vibrational modes for even number of particles. We
plotted the normal modes of  $\omega_{max}$ of cyclic structures for
N=12 ($\omega_{36} = 2183$ $cm^{-1}$) as shown in Fig.~\ref{12-13}(e) and they correspond to dipole-type of oscillations between nearest neighbor particles while for odd number of ring structures i.e. N=13 ($\omega_{39} = 2187$ cm$^{-1}$ in Fig.~\ref{12-13}(f)), similar dipole-type of oscillations between nearest neighbor particles are found but with decreasing magnitude towards opposite sides.

\begin{figure}[htb]
\centering
%  \begin{center}
    \includegraphics[width=8.6cm]{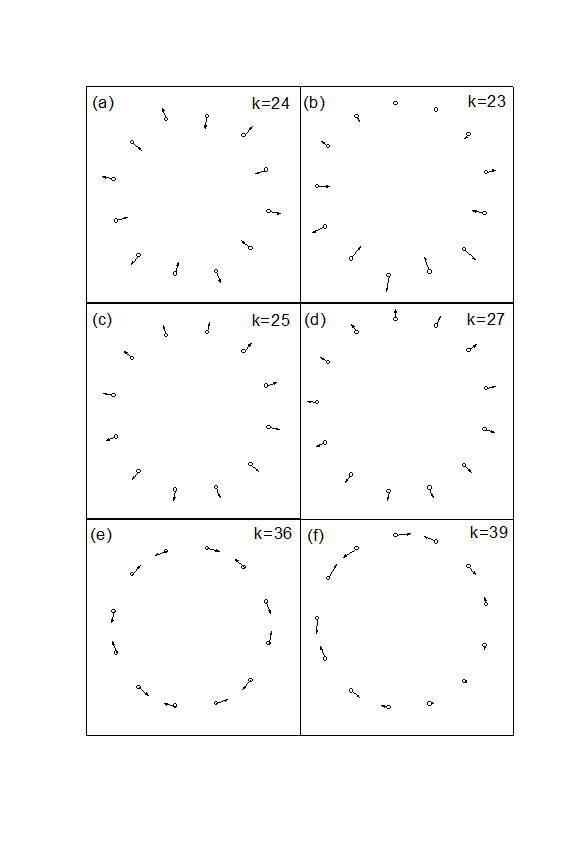}
%  \end{center}
  \caption{
  Normal modes for the out-of-phase (top panels), breathing (middle panels) and $\omega_{max}$ (bottom panels) for the  clusters N=12 (left panel) and N=13 (right panel).
}\label{12-13}
\end{figure}

\begin{figure}[htb]
\centering
%  \begin{center}
    \includegraphics[width=8.6cm]{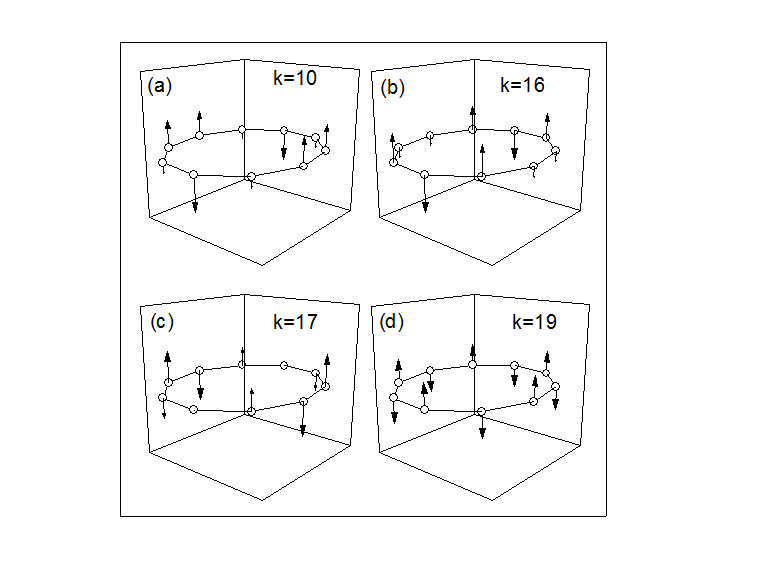}
%  \end{center}
  \caption{
  Eigenvectors for out-of-plane motion  for a cluster with N=10 particles for different mode number with frequency (a) $\omega_{10} = 165$ cm$^{-1}$, (b) $\omega_{16} = 482$ cm$^{-1}$, (c) $\omega_{17} = 482$ cm$^{-1}$ and (d) $\omega_{19} = 536$ cm$^{-1}$.
}\label{10_z}
\end{figure}

Now we investigate the eigenmodes for out-of-plane vibrations for ring clusters. For N=10 and mode k=10, the normal mode corresponds to a bending mode while for mode k=16, the normal mode corresponds to a sinusoidal type of motion. The  higher modes k=17 and k=19 have   nearest-neighbor atoms oscillating  in opposite direction perpendicular to the ring-plane with different wavelength along the ring as seen in Fig.~\ref{10_z}. For even ring structures the highest mode  for out-of-plane vibration is similar to the one shown in Fig.~\ref{10_z}(d). The out-of-plane vibration can be reasonably well fitted by  $\frac{a}{N}+b$ (solid curves in Fig. 1 for $\omega$$<$550 cm$^{-1}$) where  the fitting parameters a and b are listed in Table~\ref{table1}.
\begin{table}[tp]%
\caption{Fitting parameters for the eigenfrequencies of out-of-plane motion shown by the solid curves in Fig.1 for $\omega$$<$550 cm$^{-1}$:~$\omega=\frac{a}{N}+b$.}

\begin{tabular}{l  l}
\hline
$a$ cm$^{-1}$               &  $b$ cm$^{-1}$                  \\
                   \\
\hline
2782 $(\pm 48)$  & -107 $(\pm 5)$  \\
4258 $(\pm 120)$ & -97 $(\pm 11)$  \\
5405 $(\pm 251)$ & -62 $(\pm 20)$        \\
5686 $(\pm 59)$  & 25  $(\pm 5)$                 \\
5814 $(\pm 499)$ & 102 $(\pm 33)$        \\
4942 $(\pm 589)$ & 220 $(\pm 36)$       \\
 \hline
\end{tabular}
\label{table1}
\end{table}
 The eigenmodes belonging to  each solid curve correspond to the  same type of mode.

\begin{figure}[htb]
\centering
%  \begin{center}
    \includegraphics[width=8.6cm]{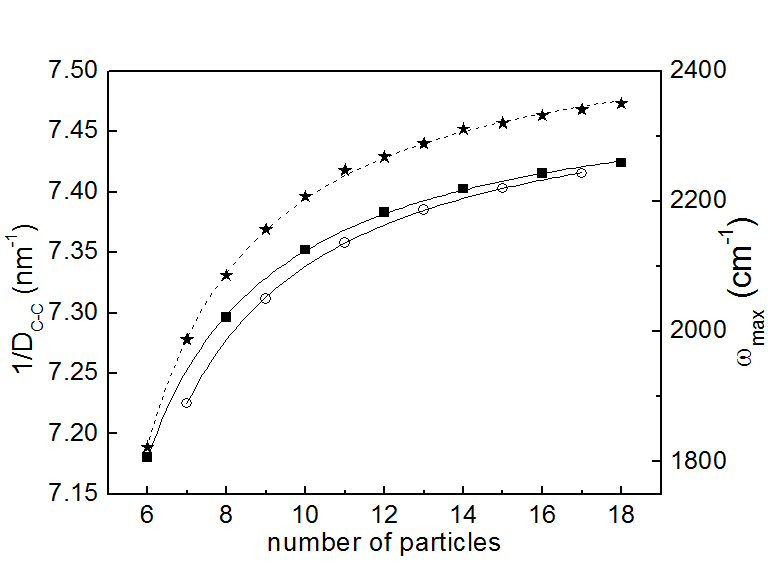}
%  \end{center}
  \caption{
$1/D_{C-C}$ (dashed curve) and $\omega_{max}$ (solid curve for even and odd N) as a function of the number of particles.
}\label{d-cc}
\end{figure}

From Fig.~\ref{fig1}, it is interesting to note that the maximum frequency in
the excitation spectrum for ring clusters, on the average, slowly increases with N. This can be explained by calculating the phonon spectrum of an infinite system. In Fig.~\ref{d-cc} we show that the minimal interparticle distance decreases
slowly with the number of C-atoms in the cluster. As a consequence, the maximum
value of the wave vector $k\approx \pi/l_0$ ($l_0$ is the mean
distance between the particles) and also the  corresponding frequency will
increase weakly with cluster size. The inverse of the inter-carbon distance could be fitted to  (dashed curve in Fig.~\ref{d-cc})
\begin{equation}
\frac{1}{D_{C-C}}=\frac{1}{D_{0}}\frac{1+aN}{1+bN}
\end{equation}
 with the fitting parameters:
$\frac{1}{D_{0}}$=7.9 nm$^{-1}$,
a =  -0.316 ($\pm$ 0.002), and
b = -0.331 ($\pm$ 0.002). We are able to fit a curve through the maximum frequency of these ring clusters using (solid curves in Fig.~\ref{d-cc})
\begin{equation}
\omega= \omega_{0}\frac{1+aN}{1+bN},
\end{equation}
where $\omega_{0}$=3200 cm$^{-1}$,  and:
i) for even N clusters
a =  -0.295 ($\pm$ 0.001), and
b = -0.395 ($\pm$ 0.002); and ii) for odd N clusters
a =  -0.276 ($\pm$ 0.001), and
b = -0.369 ($\pm$ 0.001).

%\clearpage
\end{section}
\begin{section}{NANOGRAPHENE ($N>18$)}

The clusters with $N>18$ have an inner structure and their configurations have been investigated in detail in Ref.~\onlinecite{kosi}. Their ground state configuration  can be classified in three groups: 1) nanographene clusters consisting of only hexagons, 2) clusters with pentagon on the boundary, and 3) bowl shaped configurations that have typically pentagons in the inner part of the cluster.  For the latter one  $<z^2>\neq0$, where $z$ is the position coordinate of the cluster along the out-of-plane direction and they are found for N=20, 28, 38 and 44 which are  buckled-like structures. In these clusters one  pentagon is
 surrounded by five hexagons. The normal mode oscillations in-plane and out-of-plane are now coupled, i.e. the normal mode oscillations are  3-dimensional and some of the interesting ones  are shown in Fig.~\ref{20} for N=20 and mode k=7, 9, 11 and 60. For mode k=7, the opposite atoms on the boundary of the cluster vibrate in the same direction  whereas for k=9, the normal mode corresponds to a bending mode of the cluster. But, for the higher mode k=11, the corner atoms show mixed type of oscillations while for large value of k, the  normal modes correspond to in-plane oscillations of the C-atoms.

\begin{figure}[htb]
\centering
%  \begin{center}
    \includegraphics[width=8.6cm]{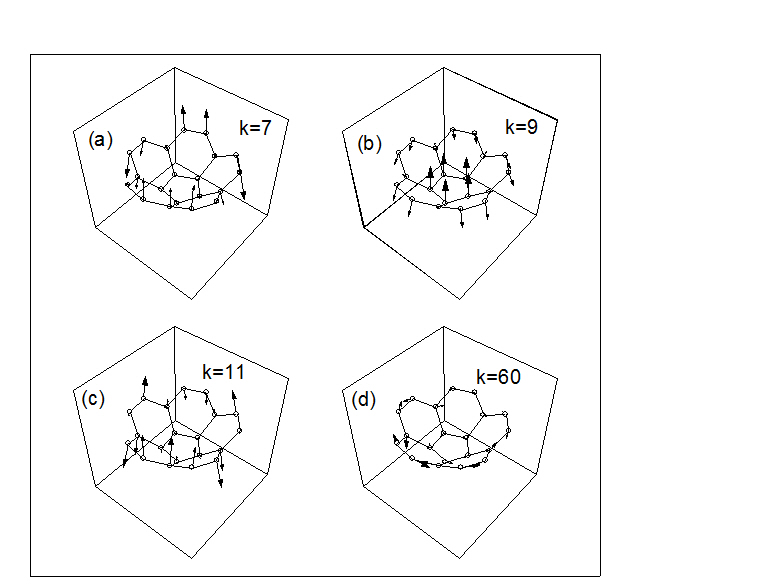}
%  \end{center}
\caption{  Eigenvectors for a cluster with N=20 particles for different mode
number with frequency (a) $\omega_{7} = 124$ cm$^{-1}$, (b) $\omega_{9} =161$ cm$^{-1}$, (c) $\omega_{11} =233 $ cm$^{-1}$ and (d) $\omega_{60} = 1772$ cm$^{-1}$.}
\label{20}
\end{figure}

%\begin{figure}[H]
%\centering
%\includegraphics[width=20cm,height=20cm]{20_k3.pdf}
%\caption{Eigenvectors for a cluster with N=20 particles for different mode
%number k=3.}
%\label{fig20k3}
%\end{figure}

\begin{figure}[htb]
\centering
%  \begin{center}
    \includegraphics[width=9.4cm]{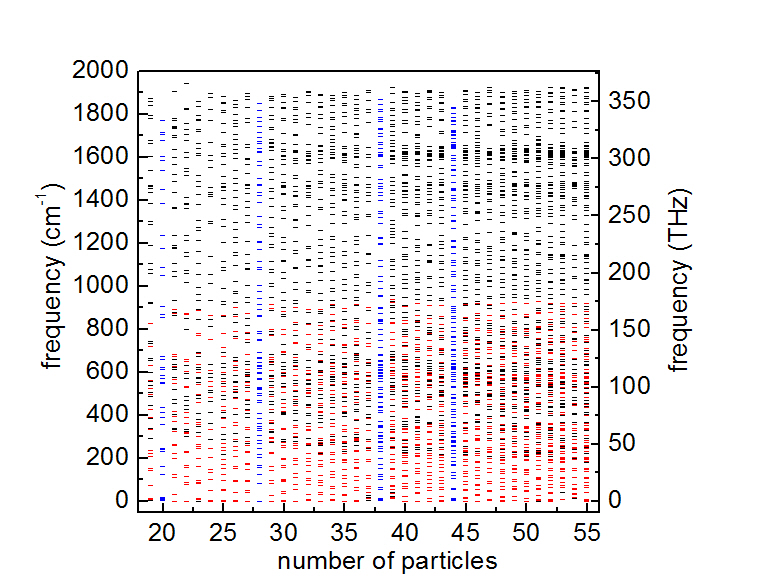}
%  \end{center}
  \caption{ (Color online)
Normal mode spectrum  as a function of the number of C atoms  for $19 \leq N \leq 55$. The modes with eigenvectors in-plane are shown in black, those with out-of-plane eigenvectors in red  and mixed type eigenvectors in blue.
}\label{cluster19-55}
\end{figure}

The normal mode frequencies for $N\geq19$ are plotted in Fig.~\ref{cluster19-55}. For N=22 and 39, a heptagon is on the boundary surrounded with pentagon and hexagon. In such a case local modes are found where only particles close to the defects (pentagon and heptagon) oscillate. If there is one pentagon on the boundary i.e. N=49, and 51, the local modes with larger amplitudes are on the opposite side of the clusters while for N=53 where a pentagon is surrounded by four hexagons, the local modes with larger amplitudes are found only near the defect. For clusters which are symmetric and without defect i.e. N=54, we found that  the local modes are situated on the boundary as shown in Fig.~\ref{cluster-defect}.
\begin{figure}[htb]
\centering
%  \begin{center}
    \includegraphics[width=8.6cm]{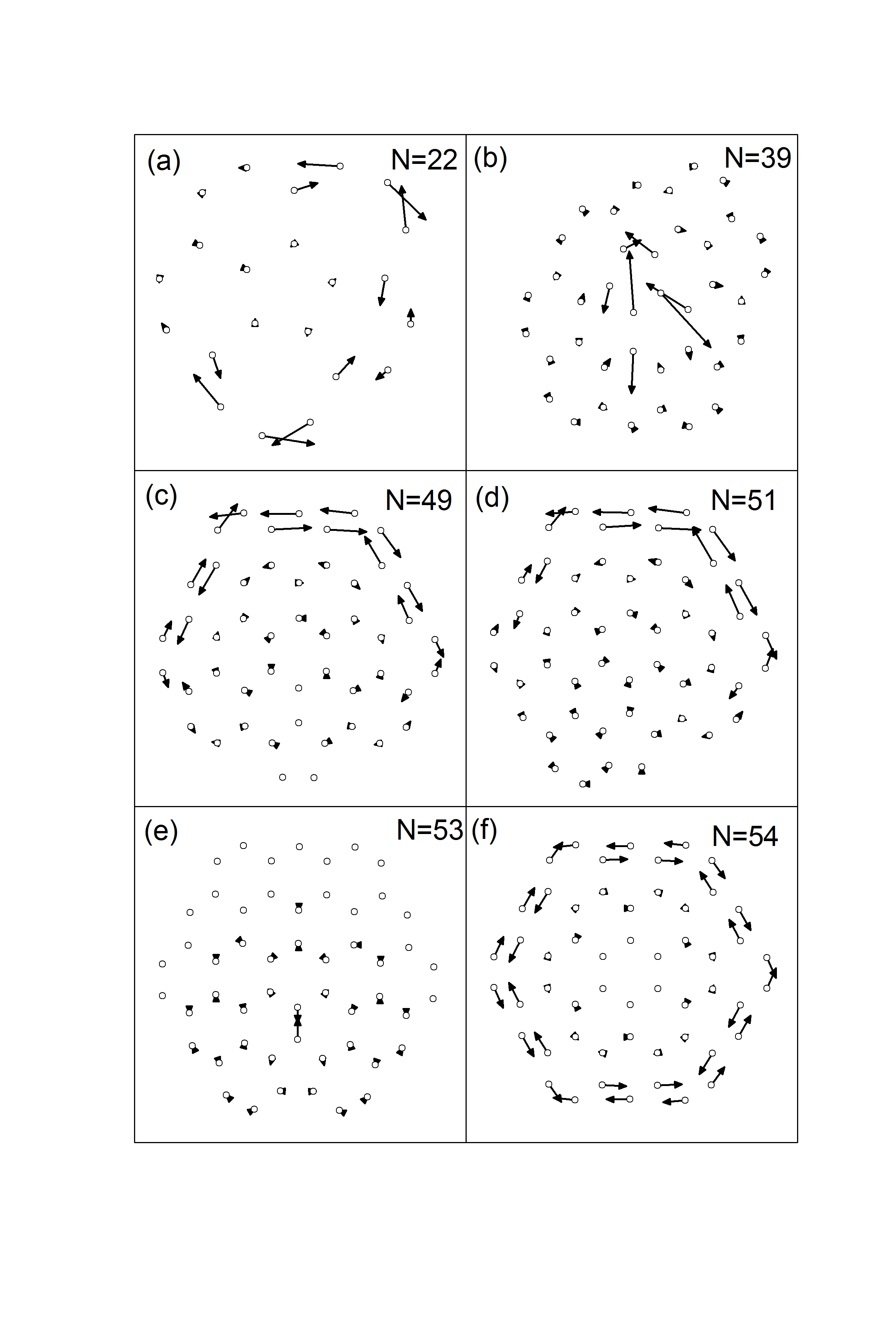}
%  \end{center}
  \caption{
Eigenvectors for highest mode number for  clusters with N=22, 39, 49,51, 53, 54 particles  with frequency (a) $\omega_{66} = 1944$ cm$^{-1}$, (b) $\omega_{117} = 1926$ cm$^{-1}$, (c) $\omega_{147} =1907 $ cm$^{-1}$, (d) $\omega_{153} = 1903$ cm$^{-1}$, (e) $\omega_{159} = 1952$ cm$^{-1}$ and (f) $\omega_{162} = 1914$ cm$^{-1}$.}\label{cluster-defect}
\end{figure}

A special subgroup of clusters consists of  hexagonal and trigonal shaped nanographene.
The clusters with size N=24 and N=54 are planar graphene structures which
are hexagonal shaped with zigzag edges. These clusters have a
close-packed structure, which  consist purely of hexagons. Let us
consider the cluster with N=54 (see Fig.~\ref{fig11}), the mode k=34  corresponds to the  breathing mode. Mode k=58 exhibits circular motion of atoms arranged in  hexagons near the 6 corners of the hexagonal disk. The next mode k=59 is similar but now hexagons are displaced by an angle of 30$^\circ$ and are centered in the middle of the sides of the hexagonal disk. Notice that for k=58 the 6 rotational oscillations are in the same direction while for k=59 they alternate, i.e. vortex/anti-vortex like arrangements. For
higher modes k=139, only the inner particles participate in the normal mode oscillations. For k=148, a dipole type of oscillations is found of nearest neighbor C-atoms arranged in a shell around the middle between the center and the perimeter. For higher frequency only the outer particles oscillate  while the inner particles exhibit only very small
displacements as shown in Fig.~\ref{fig11}.

\begin{figure}[htb]
\centering
%  \begin{center}
    \includegraphics[width=8.6cm]{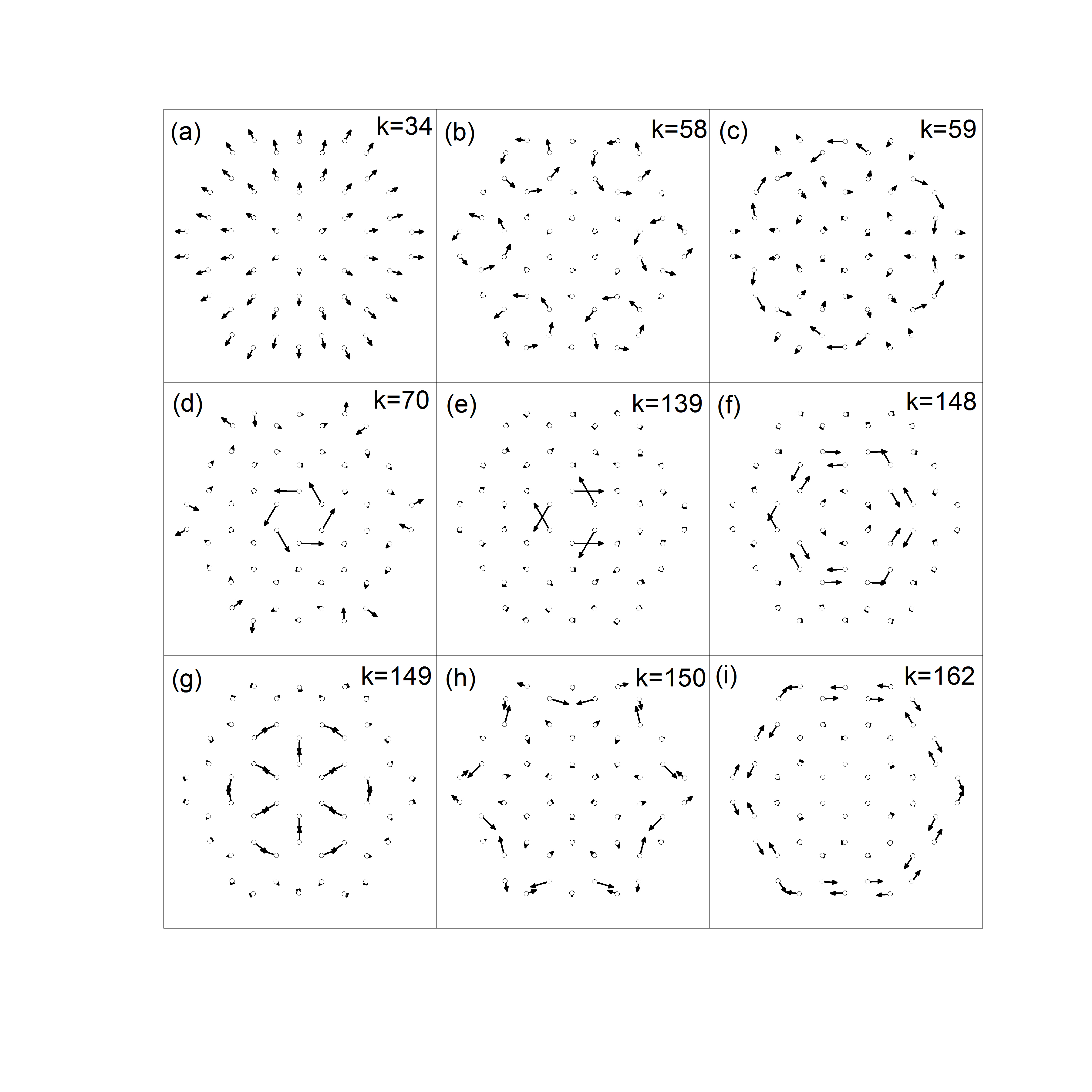}
%  \end{center}
  \caption{
Eigenvectors for a cluster with N=54 particles for different mode
number with frequency (a) $\omega_{34} = 313$ cm$^{-1}$, (b) $\omega_{58} = 531$ cm$^{-1}$, (c) $\omega_{59} =534 $ cm$^{-1}$, (d) $\omega_{70} = 593$ cm$^{-1}$, (e) $\omega_{139} = 1589$ cm$^{-1}$,  (f) $\omega_{148} = 1624$ cm$^{-1}$, (g) $\omega_{149} = 1628$ cm$^{-1}$,  (h) $\omega_{150} = 1638$ cm$^{-1}$ and (i) $\omega_{162} = 1914$ cm$^{-1}$.}\label{fig11}
\end{figure}

From Fig.~\ref{cluster19-55} we notice that there is a region along the frequency axis with a higher density of normal modes. Examples of the displacements of such modes are shown in Fig.~\ref{fig12} for N=54.  Mode k=27 corresponds to the excitation of a vortex/antivortex pair. Mode k=33 consists of rotational oscillations around the hexagonal corners of the nanodisk, while for k=35 these rotational motions are centered around the middle of the hexagonal sides. Notice that mode k=45 consists of a central large vortex motion surrounded by 6 anti-vortex type of motions situated closer to the edge of the nanodisk.
\begin{figure}[htb]
\centering
%  \begin{center}
    \includegraphics[width=8.6cm]{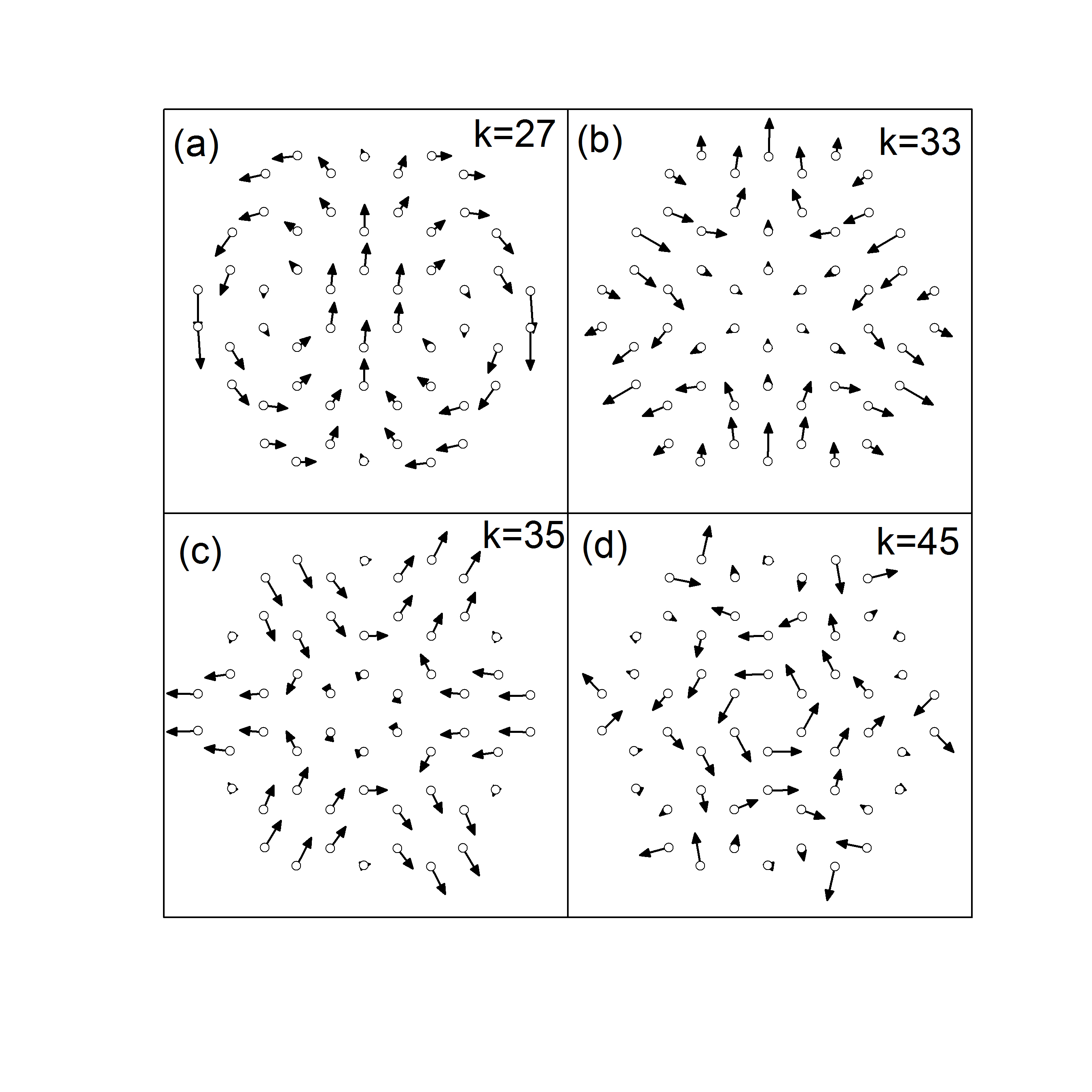}
%  \end{center}
  \caption{
 Eigenvectors for the cluster with N=54 particles for four different values of the mode number k with frequency (a) $\omega_{27} = 244$ cm$^{-1}$, (b) $\omega_{33} =302$ cm$^{-1}$, (c) $\omega_{35} =314 $ cm$^{-1}$ and (d) $\omega_{45} = 408$ cm$^{-1}$.
}\label{fig12}
\end{figure}

The nanodisk  cluster with  N=46 carbon atoms has a metastable configuration consisting of closed zigzag edges and a trigonal-shaped structure~\cite{kosi}. As shown in Fig.~\ref{fig13} the mode k=18 consists of an asymmetric vortex/antivortex pair while mode k=29 is a clear breathing mode. The modes k=24 and k=32 show three rotations situated close to the three corners of the trigonal nanodisk, but notice that the rotation direction and the position of the center of the rotation is not always the same for both modes .
\begin{figure}[htb]
\centering
%  \begin{center}
    \includegraphics[width=8.6cm]{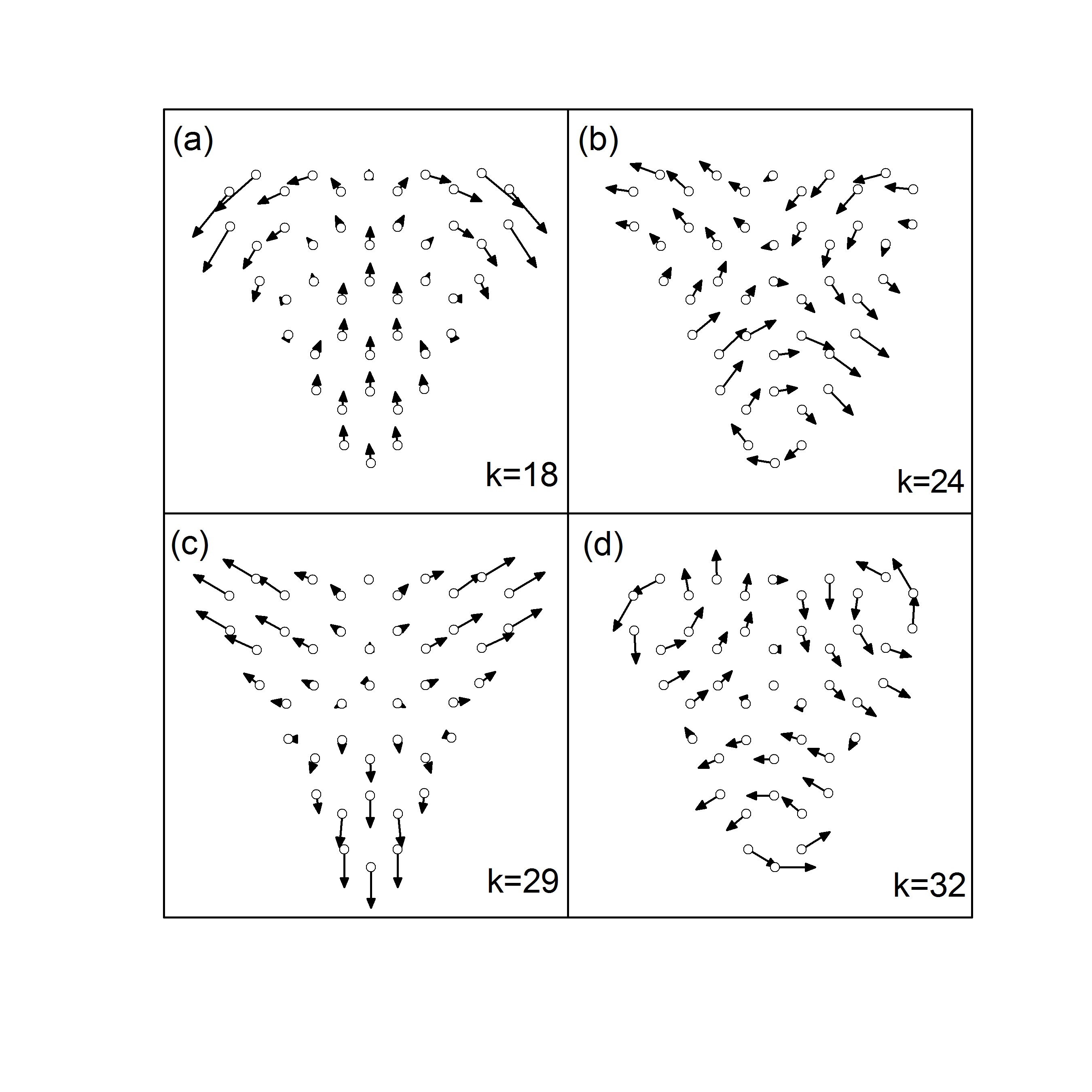}
%  \end{center}
  \caption{
Eigenvectors for a cluster with N=46 particles for different mode
number with frequency (a) $\omega_{18} = 185$ cm$^{-1}$, (b) $\omega_{24} =244$ cm$^{-1}$, (c) $\omega_{29} =288 $ cm$^{-1}$ and (d) $\omega_{32} = 322$ cm$^{-1}$. }\label{fig13}
\end{figure}

We  plotted the average distance  between the C-C atoms as a function of the number of particles N. The average radius  increases linearly with the chain length for (N=3-5) which could be fitted to $D_{C-C}=D_{0}+a \times N$ where $D_{0}=0.13$ nm and a=0.00055 $(\pm$ 0.00009) nm (see solid curve in Fig.~\ref{d_ccvsN}). For ring clusters it decreases exponentially as $D_{C-C}=D_{0}+a \times exp(b \times N)$ (see solid curve in Fig.~\ref{d_ccvsN}) where the parameter $D_{0}=0.134$ nm, $a=0.039 (\pm 0.001)$ nm and $b=-0.337 (\pm 0.007)$ as the number of particles increases. For $N>18$ the average C-C distance fluctuates as function of N around the average value 0.141 nm which compares with 0.142 nm for the C-C distance in bulk graphene. The clusters with pentagon and Stone-Wales defects inside the clusters show larger average C-C distance than the  pure hexagonal structures while the clusters with pentagons on the boundary (i.e. N=26, 31, 34, 36, 41, 46, 51 and 53) show the highest average distance as shown in Fig.~\ref{d_ccvsN}.
\begin{figure}[htb]
\centering
%  \begin{center}
    \includegraphics[width=8.6cm]{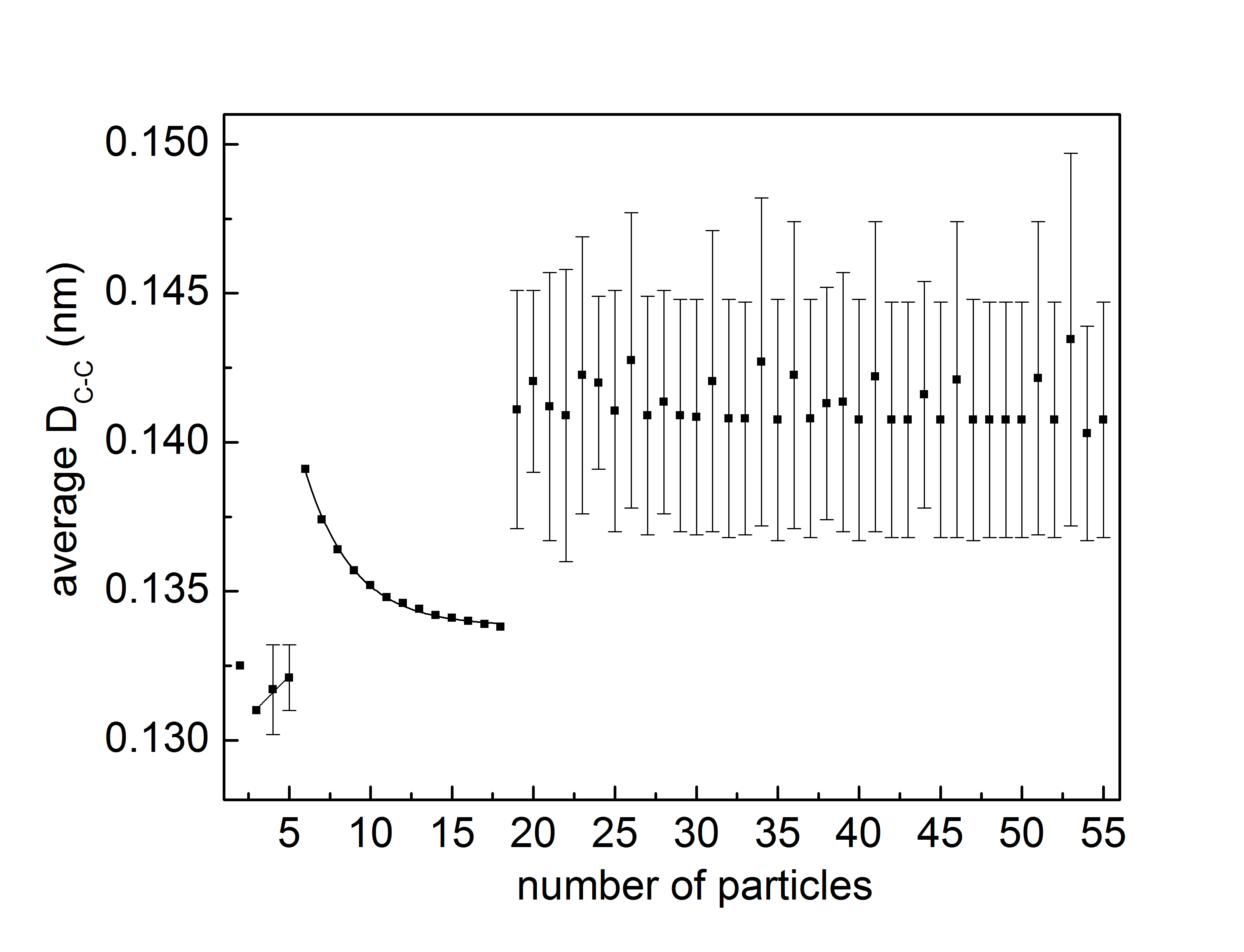}
%  \end{center}
  \caption{
The average distance between C-C atoms as a function of the number of particles in the graphene cluster. The error bars indicate the range of C-C distances within each cluster.}\label{d_ccvsN}
\end{figure}

\begin{figure}[htb]
\centering
%  \begin{center}
    \includegraphics[width=9.4cm]{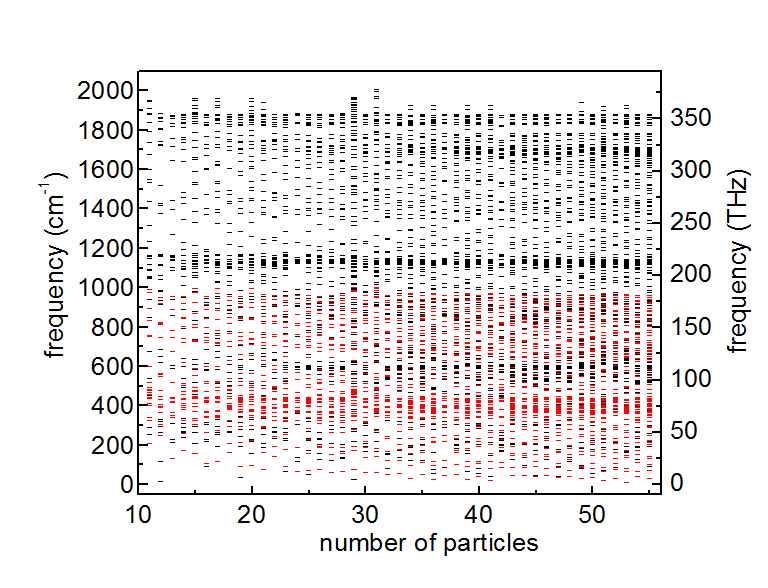}
%  \end{center}
  \caption{ (Color online)
 Normal mode spectrum  as a function of the number of C particles  for $11 \leq N \leq 55$ H-passivated clusters. The modes with eigenvectors in plane are shown in black, those having out-of plane eigenvectors in red.
}\label{fig_h_pass}
\end{figure}

For completeness we also plotted (in Fig.~\ref{fig_h_pass}) the energy  spectrum of the normal modes for ground-state H-passivated $C_{N}$ (N=11-55) clusters~\cite{kosi}. Notice that different from Fig.~\ref{cluster19-55} (i.e. non H-passivated graphene) there is a  region with enhanced density of modes, i.e. around 2900 $cm^{-1}$. The corresponding modes are connected to oscillations of the C-H atoms. This is illustrated in Fig.~\ref{h-pass} for  H-passivated nanographene clusters with N=53 and 54. In the optical region which is a region along the frequency axis with higher density of normal modes, for N=53 with k=172  the corresponding local normal modes with larger amplitudes are on the opposite side of the defect while for N=54 with k=187, only the inner C-atoms oscillate with larger amplitude (Figs.~\ref{h-pass} (a, b)). The highest frequency mode of the N=53 and 54 clusters (Figs.~\ref{h-pass} (c, d)) is a local mode  near the local defect where mainly outer C-H atoms of the armchair C-atoms  oscillate.

\begin{figure}[htb]
\centering
%  \begin{center}
    \includegraphics[width=8.6cm]{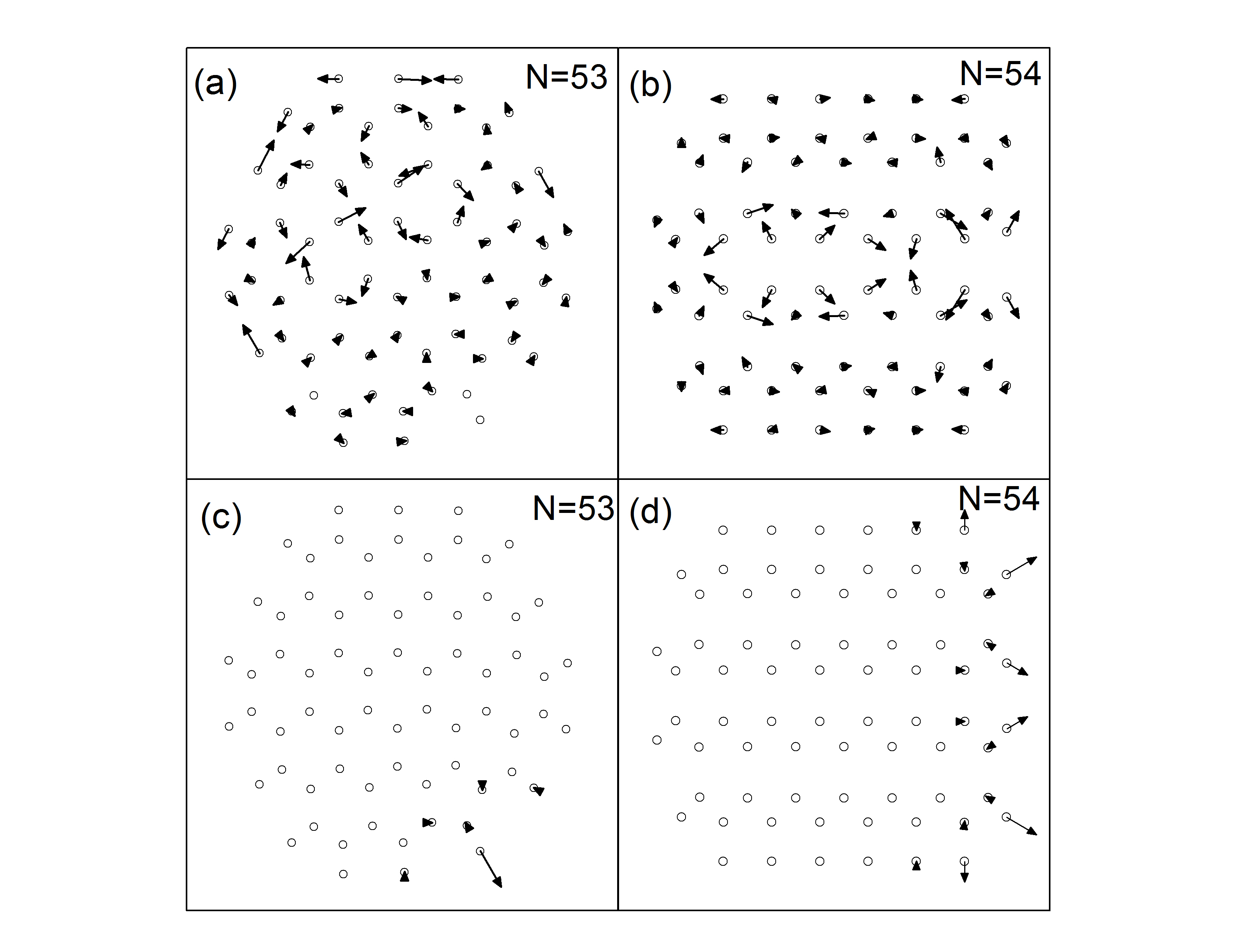}
%  \end{center}
  \caption{
Eigenvectors for a cluster with N=53, and 54 with H-passivation for  k=172 for N=53 and k=187 for N=54, and for $\omega_{max}$ corresponding to frequency (a) $\omega_{172} = 1693$ cm$^{-1}$, (b) $\omega_{187} =1698$ cm$^{-1}$, (c) $\omega_{210} =2925 $ cm$^{-1}$ and (d) $\omega_{222} = 2915$ cm$^{-1}$.}\label{h-pass}
\end{figure}

%\clearpage
\end{section}
\begin{section}{PHONON DENSITY OF STATES}

 In order to compare the normal modes with those of graphene we calculate the phonon density of states (PDOS). Furthermore, we calculated the PDOS of exactly trigonal and hexagonal  two dimensional and defective clusters and analyze what is the effect of defects on the density of states. Because of the discreteness of the normal modes frequency spectrum we introduced a Gaussian broadening
\begin{equation}
\rho(\omega) = \sum_{i=1}^{3N}exp{(-(\omega-\omega_{i})^{2}/\delta \omega^{2})},\\\\
\end{equation}
where the summation is over all normal modes, $\omega_{i}$ is the normal mode frequency of the $\textit{i}^{th}$ mode, broadening is chosen  $\delta \omega =30$  cm$^{-1}$ and N is the total number of atoms in the cluster.  We notice  that the introduction of a pentagon in the N=53 cluster as compared to the perfect  hexagon lattice in N=54 introduces more high frequency modes as shown in Fig.~\ref{pdos_53} which are due to modes localized around the defect.

\begin{figure}[htb]
\centering
%  \begin{center}
    \includegraphics[width=8.6cm]{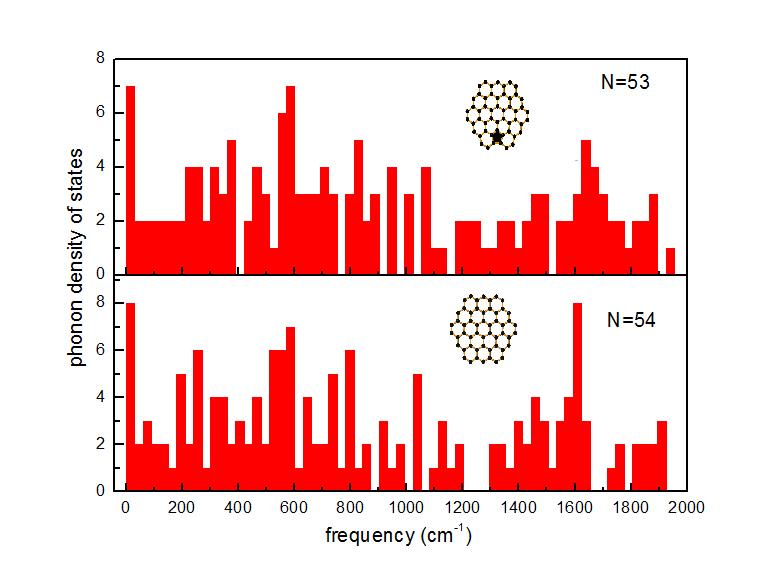}
%  \end{center}
  \caption{ (Color online)
The phonon density of states for clusters with N= 53 and 54 carbon atoms.
}\label{pdos_53}
\end{figure}

\begin{figure}[htb]
\centering
%  \begin{center}
    \includegraphics[width=8.6cm]{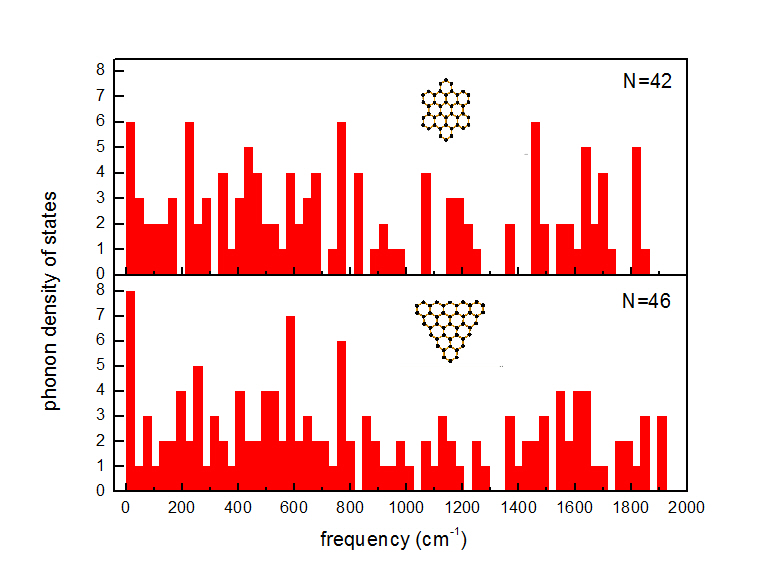}
%  \end{center}
  \caption{ (Color online)
The phonon density of states for clusters with N=42 (hexagonal with armchair) and 46 (trigonal) carbon atoms.
}\label{pdos_42}
\end{figure}
Fig.~\ref{pdos_42} shows the phonon density of states (PDOS) of armchair hexagonal and zigzag trigonal clusters. Notice that several of the peaks in the PDOS of the hexagonal clusters are split into two peaks in the PDOS of the trigonal shaped cluster. This is a consequence of the reduced rotational symmetry of the trigonal cluster.

We compare in Fig.~\ref{pdos_55} the phonon density of states of clusters N=55 and 1600 with those for graphene~\cite{Vandescuren} which shows nicely the convergence of the phonon spectrum of nanographene to bulk graphene. For N=1600 we almost recover the  theoretical results of graphene~\cite{Vandescuren}. Notice that for small clusters there is a pronounced PDOS for $\omega=0$ which is due to the relative importance of the $\omega=0$ translational and rotational motion.
\begin{figure}[htb]
\centering
%  \begin{center}
    \includegraphics[width=8.6cm,height=8cm]{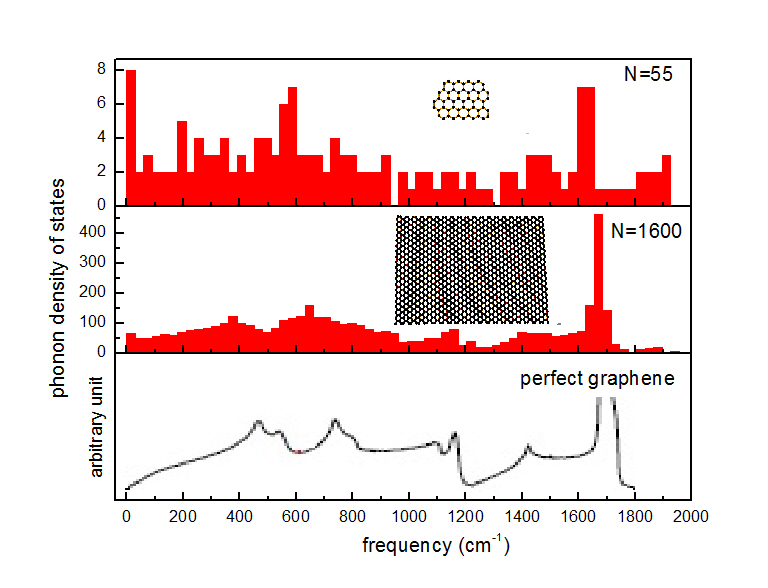}
%  \end{center}
  \caption{ (Color online)
The phonon density of states for clusters with N=55, 1600 and perfect graphene (bottom panel is taken from Ref.~\onlinecite{Vandescuren}).
}\label{pdos_55}
\end{figure}

  The effect of H-passivation of the edge atoms of the cluster is shown in Fig.~\ref{pdos_53_h} for N=53 and 55. The C-H bonds  introduce high frequency modes  in the N=53 and N=55 clusters as compared to the non H-passivated N=53 and N=55 clusters.  The H-passivated clusters exhibit pronounced peaks around ~1700 and ~2900 cm$^{-1}$ which correspond to the optical region and the C-H atom frequency region, respectively. We also notice  that the introduction of a pentagon in the N=53 cluster as compared to the perfect  hexagon lattice for N=55 introduces more high frequency modes as shown in Fig.~\ref{pdos_53_h} similar to the case for  non H-passivated clusters (see Fig.~\ref{pdos_53}) which are due to modes localized around the defect.

\begin{figure}[htb]
\centering
%  \begin{center}
    \includegraphics[width=8.6cm]{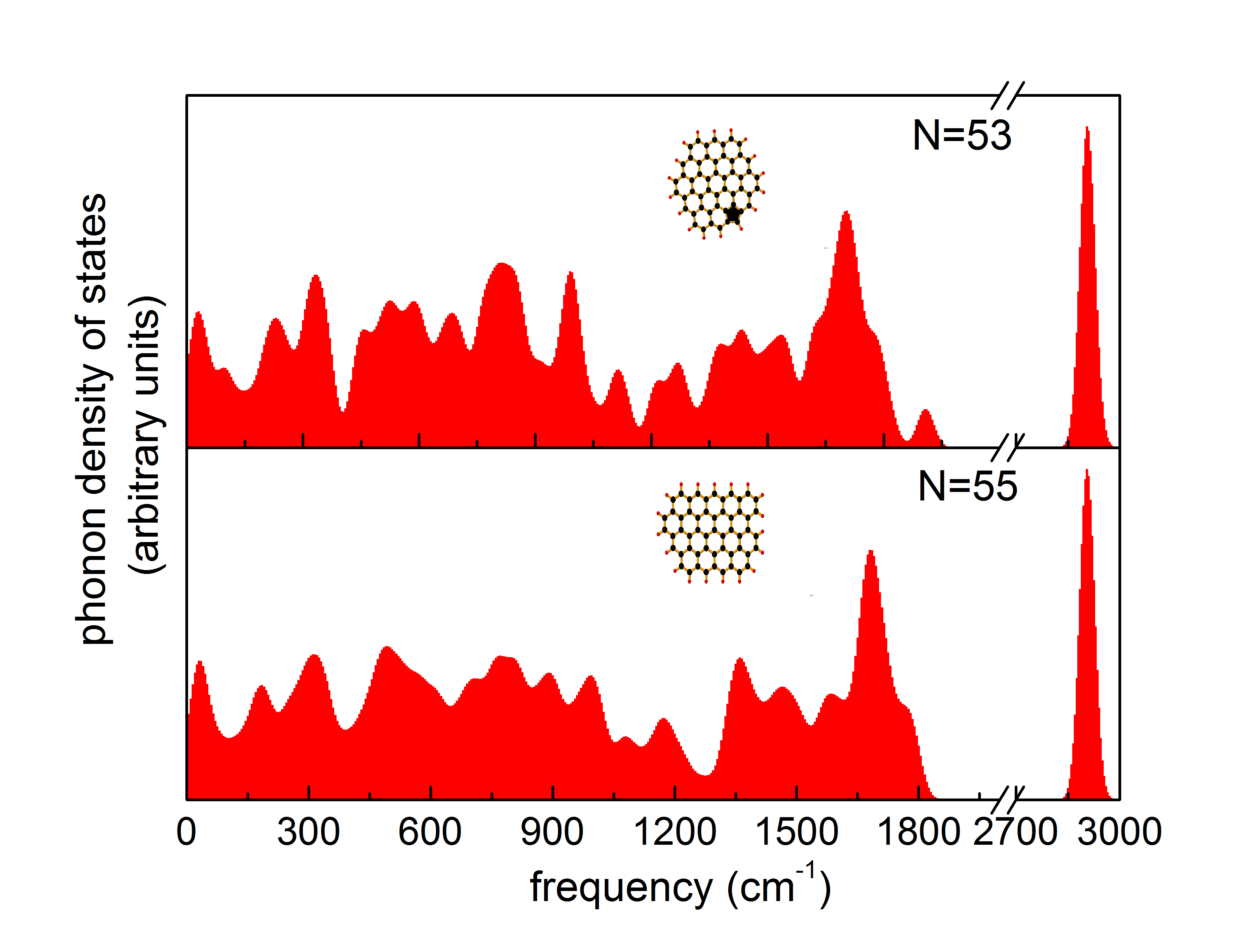}
%  \end{center}
  \caption{ (Color online)
The phonon density of states for H-passivated clusters with N=53 and 55.
}\label{pdos_53_h}
\end{figure}

The modes can have a shear-like or a compression-like character. The compressional and shear properties can be extracted from the divergence and rotation of the velocity field. Here, we will associate a single number for the shear-like and compression-like character of the different modes by calculating the squared average of the divergence $\vec{\nabla}.\vec{\upsilon}$ and the vorticity $(\vec{\nabla} \times\vec{\upsilon})_{z}$ of the velocity field, following the approach of Ref.~\onlinecite{Schweigert}.

The z component of the rotation $\Psi_{r}(k)=\vec{e}_{z}.rot\Psi(k)$ and the divergence $\Psi_{d}(k)=div\Psi(k)$ of the field of the eigenvectors of mode k are
 \begin{equation}
\Psi_{d}(k)=\frac{1}{N}\sum_{i=1}^{N} \Psi_{d,i}^{2}(k),
\end{equation}
\begin{equation}
\Psi_{r}(k)=\frac{1}{N}\sum_{i=1}^{N} \Psi_{r,i}^{2}(k),
\end{equation}
where the values of $\Psi_{d,i}(k)$ and $\Psi_{r,i}(k)$ for the $\textit{i}^{th}$ particle are, respectively, given by
\begin{equation}
\Psi_{d,i}(k)=\frac{1}{J}\sum_{j=1}^{J} (\vec{r}_{i}-\vec{r}_{j}) . [\vec{A}_{i}(k)-\vec{A}_{j}(k)]/|\vec{r}_{i}-\vec{r}_{j}|^{2},
\end{equation}

\begin{equation}
\Psi_{r,i}(k)=\frac{1}{J}\sum_{j=1}^{J} (\vec{r}_{i}-\vec{r}_{j}) \times [\vec{A}_{i}(k)-\vec{A}_{j}(k)]/|\vec{r}_{i}-\vec{r}_{j}|^{2}.
\end{equation}
Here, j and J denote the index and the number of neighboring particles of particle i, respectively.  $\vec{r}_{j}$ is the positional coordinate of a neighboring particle and $\vec{A}_{i}(k)$ is the eigenvector of particle i for mode k. Note also that we calculate the squared average over all the particles because the simple spatial average is of course zero.

We plot $\Psi_{d}(k)$ and $\Psi_{r}(k)$ as a function of the mode number k for cluster N=39 which contains a  5-7 defect, cluster N=49 with pentagon defect on the boundary, the  pure symmetric hexagonal cluster with N=54 and for a large cluster having N=398 particles. In general, the lower eigenfrequency spectrum corresponds to  rotational type of excitations which are vortex-antivortex like excitations for large clusters (see Fig.~\ref{fig12}). Vortex excitation is only expected for sufficiently large clusters, because the velocity of dissipation of the vortex energy is inversely proportional to $R^{2}$, where $R$ is the radius, which increases with increasing cluster size.  In the second half of the spectrum the divergence $\Psi_{d}(k)$, which corresponds to compression-like modes, is appreciably different from zero and we have  mixed modes that have  both a shear-like and a compression-like component. For symmetric clusters and defective clusters with pentagon on the boundary, the highest modes consist of only  shear-like modes (see Figs.~\ref{cluster-defect}(c, d, f)). For  cluster N=39 with a 5-7 defect, compression-like modes appear for the highest frequency modes (see Fig.~\ref{cluster-defect}(b)).

\begin{figure}[htb]
\centering
%  \begin{center}
    \includegraphics[width=8.6cm,height=8.6cm]{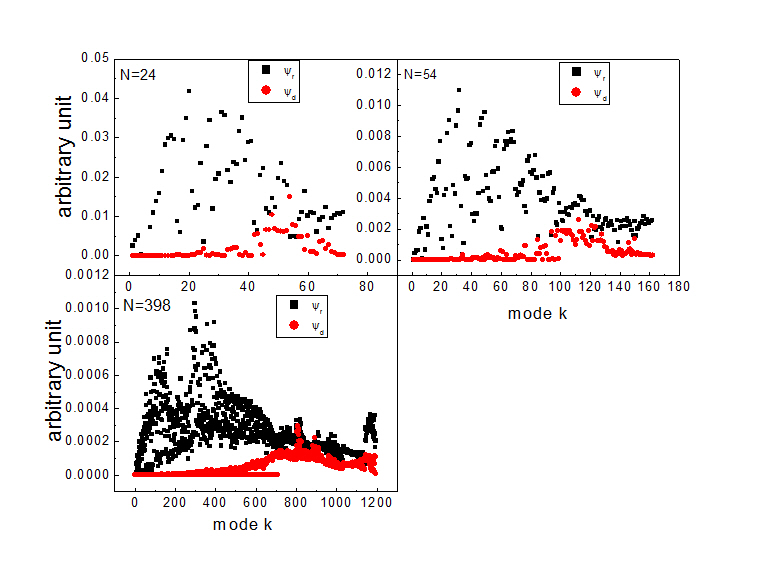}
%  \end{center}
  \caption{ (Color online)
Distribution of the rotation $\Psi_{r}$ and the divergence $\Psi_{d}$ as function of the mode number k for N=39, 49, 54 and 398 particles.
}\label{rotor}
\end{figure}

%\clearpage

The heat content of a system is directly related to the specific heat whose lattice contribution is determined by the phonons i.e. the eigenmodes of the system, while electronic contribution to it can be neglected even at a few Kelvins \cite{Benedict}. The specific heat C$_{\text P}$ per unit mass, at temperature T is within the harmonic approximation \cite{Maradudin} given by the following expression:
\begin{equation}
C_P = \frac{k_B}{MN}\sum_{k=1}^{3N}\left(\frac{\hbar \omega_{k}}{2k_{B}T}\right)^{2} csch^{2}\left(\frac{\hbar \omega_{k}}{2k_{B}T}\right) ,\\\\
\end{equation}
where the summation is over all normal modes, k$_{\text B}$ is the Boltzmann constant, M is the mass of each carbon atom and N is the total number of atoms in the cluster. From Eq. (15), the specific heat depends sensitively on the characteristics of the phonon spectrum and on its vibrational density of states.

The numerical results for N=5, 18, 54 and 398 are shown in Fig.~\ref{specific_low}. It may be noted that at low temperature, cluster sizes with N=5, 18, 54 and 398 have a specific heat C$_{\text P}$ which exhibits a clear linear dependence in T. The slope of the low temperature C$_{\text P}$ behavior is shown in the inset of Fig.~\ref{specific_low} as a function of N which can be fitted to: a + b $\times$ exp(c $\times$ N) (solid curve) where a=3.77 J kg$^{-1}$ K$^{-1}$, b=3.87 $(\pm 1.23)$ J kg$^{-1}$ K$^{-1}$ and c=-0.05  $(\pm 0.01)$ K$^{-1}$.

\begin{figure}[htb]
\centering
%  \begin{center}
    \includegraphics[width=8.6cm]{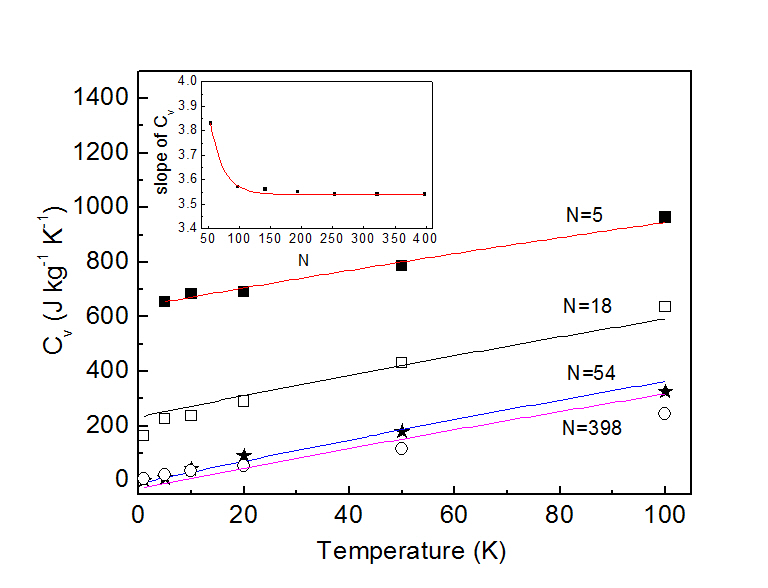}
%  \end{center}
  \caption{ (Color online)
Phonon contribution to the constant pressure heat capacity of clusters with  N=5, 18, 54 and 398 at low temperature. The curves are linear fits to the numerical results. In the inset we show the N-dependence of the slope of the low temperature linear T-dependence of the specific heat.
}\label{specific_low}
\end{figure}

The high temperature behavior is shown in Fig.~\ref{specific_high}. For large temperatures all results approach the value 2078 J kg$^{-1}$ K$^{-1}$ which is in good agreement with the bulk graphene value given in Refs.~\cite{Mounet, janina}.

We fitted the numerical results using the formula
\begin{equation}
C_{P}= a + b \times exp(c \times T),
\end{equation}
where  a=2078.0 J kg$^{-1}$ K$^{-1}$ and b and c are  fitting parameters.
After fitting the curve, we obtained  the  parameters listed in Table~\ref{table2}. Notice that $\lvert b\rvert$ increases with N while $\lvert c\rvert$ decreases.
\begin{table}[tp]%
\caption{Fitting parameters for the specific heat (Eq. (16))}
\begin{tabular}{l l l}
\hline
N              &  b (J kg$^{-1}$ K$^{-1})$  & c (K$^{-1})$          \\

\hline
 5               &  -1502 $(\pm 38)$          &-0.00237  $(\pm 0.00021)$   \\
 18              &  -1865 $(\pm 21)$          &-0.00236  $(\pm 0.00010)$   \\
 54              &  -2087  $(\pm 10)$         &-0.00195  $(\pm 0.00003)$   \\
 398             &  -2109 $(\pm 14)$          &-0.00178  $(\pm 0.00004)$    \\
 \hline
\end{tabular}
\label{table2}
\end{table}

\begin{figure}[htb]
\centering
%  \begin{center}
    \includegraphics[width=8.6cm]{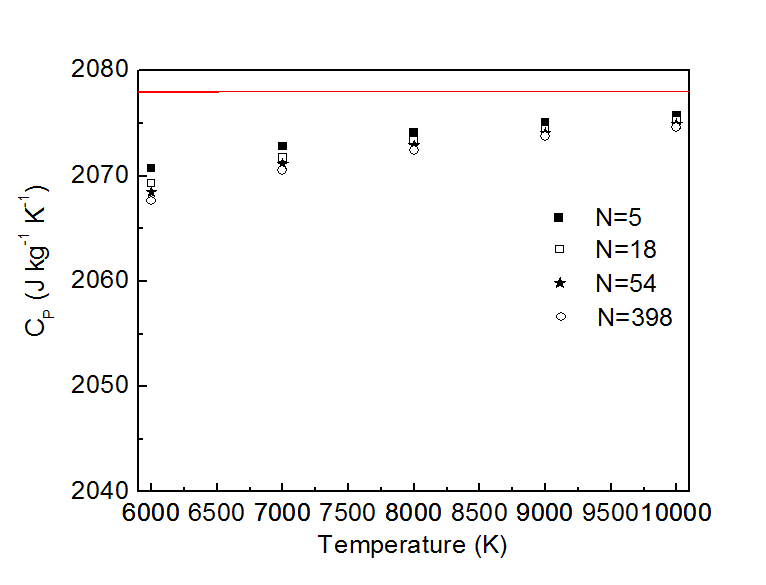}
%  \end{center}
  \caption{ (Color online)
Phonon contribution to the constant pressure heat capacity of clusters with  N=5, 18, 54 and 398 at high temperature.  The numerical results approach the fitted value 2078 $J kg^{-1} K^{-1}$.
}\label{specific_high}
\end{figure}

\end{section}
\begin{section}{Summary}

Atomistic simulations were performed using the Brenner second-generation reactive bond order (REBO) interatomic potential function in order to study the vibrational properties of nanographene. In the present work we investigated the frequency of the normal modes as a function of the number of particles for nanographene $C_{N}$ ($2\leq N \leq55$)
and H-passivated carbon clusters ($11\leq N \leq55$). We also presented a simple model for  the eigenfrequencies for linear chain and ring structures and obtained analytical results that are in good agreement with our
numerical results. The phonon density of states of different clusters are compared with theoretical results for graphene and found to be in  good agreement for the acoustic and optical phonon density in the case of large clusters. Using the eigenfrequencies of the normal modes we calculated the specific heat within the harmonic approximation for different size carbon clusters and found that  in the  high temperature limit they approach the bulk value as given in Refs. \cite{Mounet, janina}.
\end{section}

\section*{Acknowledgments}

This work was supported by the ESF-EuroGRAPHENE project CONGRAN and the Flemish Science Foundation (FWO-Vl).

\bibliographystyle{plainnat}

\end{document}